\documentclass[prd,nofootinbib,preprint,superscriptaddress,twocolumn,10pt]{revtex4}
\pdfoutput=1

\usepackage{amsmath,amssymb}
\usepackage{epsfig}
\usepackage{graphicx}
\usepackage[usenames,dvipsnames]{color}
\usepackage{subfigure}
\usepackage{slashed}
\usepackage{comment}
\usepackage[colorlinks,citecolor=blue]{hyperref}
\usepackage{color}
\newcommand{\lsim}{\mbox{\raisebox{-.6ex}{~$\stackrel{<}{\sim}$~}}}
\newcommand{\gsim}{\mbox{\raisebox{-.6ex}{~$\stackrel{>}{\sim}$~}}}

\begin{document}
	\title{A new realisation of light thermal self-interacting dark matter and detection prospects}

	\author{Debasish Borah}
	\affiliation{Department of Physics, Indian Institute of Technology Guwahati, Assam 781039, India}
	\author{Satyabrata Mahapatra}
	\affiliation{Department of Physics, Indian Institute of Technology Hyderabad, Kandi, Sangareddy 502285, Telangana, India}
	\author{Narendra Sahu}
	\affiliation{Department of Physics, Indian Institute of Technology Hyderabad, Kandi, Sangareddy 502285, Telangana, India}
	
	\begin{abstract}
Light dark matter (DM) in the GeV ballpark faces weaker constraint from direct detection experiments. If such DM also has sufficient self-interactions due to a light mediator, it can alleviate the small-scale problems of the cold dark matter (CDM) paradigm while being consistent with the latter at large scales, as suggested by astrophysical observations. However, generating correct thermal relic for such light DM is challenging. While GeV scale CDM typically leads to thermal overproduction due to the absence of sufficient annihilation channels, self-interacting dark matter (SIDM) typically has under-abundant thermal relic due to its large annihilation rates into mediator particles. In this letter, we propose a minimal realisation of GeV scale SIDM with correct thermal relic, where one 
of the three singlet fermions, responsible for seesaw origin of light neutrino masses, assists in generating correct SIDM relic 
density via freeze-out. 
The setup thereby encompasses astrophysical as well as cosmological aspects of light DM while simultaneously providing a solution to the origin of light neutrino mass. Due to such connection to several frontiers, there 
exist exciting discovery prospects in terms of direct and indirect search of DM, laboratory and cosmology based dark photon signatures and effective light neutrino mass.

	\end{abstract}	
	\maketitle
	
\noindent
{\bf Introduction:} Self-interacting dark matter (SIDM) has emerged as an alternative to the standard cold dark matter (CDM) hypothesis due to the small-scale issues 
	like too-big-to-fail, missing satellite and core-cusp problems faced by the latter \cite{Spergel:1999mh, Tulin:2017ara, Bullock:2017xww}. While CDM is collisionless, SIDM scenario requires sizeable self-interaction, parametrised in terms of cross-section to mass ratio as $\sigma/m \sim 1 \; {\rm cm}^2/{\rm g} \approx 2 \times 10^{-24} \; {\rm cm}^2/{\rm GeV}$ \cite{Buckley:2009in, Feng:2009hw, Feng:2009mn, Loeb:2010gj, Zavala:2012us, Vogelsberger:2012ku}. A DM with a light mediator can not only give rise to such a large self-interaction but also to velocity dependent DM self-interactions which solves the small-scale issues while being consistent with standard CDM properties at large scales \cite{Buckley:2009in, Feng:2009hw, Feng:2009mn, Loeb:2010gj, Bringmann:2016din, Kaplinghat:2015aga, Aarssen:2012fx, Tulin:2013teo}. However, the same coupling of DM with light mediators leads to large DM annihilation rates, typically generating under-abundant relic in low DM mass regime of a few GeV.
	
	Light thermal DM regime $(M_{\rm DM} \lsim \mathcal{O}(10 \, \rm GeV))$ has received lots of attention in recent times, particularly due to weaker constraints from direct detection experiments \cite{LUX-ZEPLIN:2022qhg}. However, it is challenging to get correct relic of thermal DM in low mass regime, typically due to insufficient annihilation cross-section. For DM interactions typically in the weakly interacting massive particle (WIMP) ballpark, the requirement of DM not overclosing the universe leads to a lower bound on its mass, around a few GeV \cite{Lee:1977ua, Kolb:1985nn}\footnote{This bound, derived for fermionic DM candidates, can be different for scalar DM \cite{Boehm:2003hm}.}. However, with inclusion of light mediators or additional particles it is possible to realise light thermal DM, as pointed out in several works \cite{Pospelov:2007mp, DAgnolo:2015ujb, Berlin:2017ftj, DAgnolo:2020mpt, Herms:2022nhd, Jaramillo:2022mos}. Inclusion of such additional fields prevent such light thermal DM from being overproduced. On the other hand, for light thermal SIDM, the relic is underproduced due to too large annihilation rates into light mediators. Despite the fact that there are several production mechanisms for SIDM in the literature~\cite{Kouvaris:2014uoa, Bernal:2015ova, Kainulainen:2015sva, Hambye:2019tjt, Cirelli:2016rnw, Kahlhoefer:2017umn, Belanger:2011ww}, finding a correct thermal relic of GeV scale SIDM is a challenging task which needs to be addressed in a beyond standard model (BSM) physics scenario. While pure thermal relic is challenging, a hybrid setup of thermal and non-thermal contribution can lead to correct SIDM relic density as studied in refs. \cite{Dutta:2021wbn, Borah:2021pet, Borah:2021rbx, Borah:2021qmi}.

\begin{figure}[h!]
		\centering
		\includegraphics[scale=0.4]{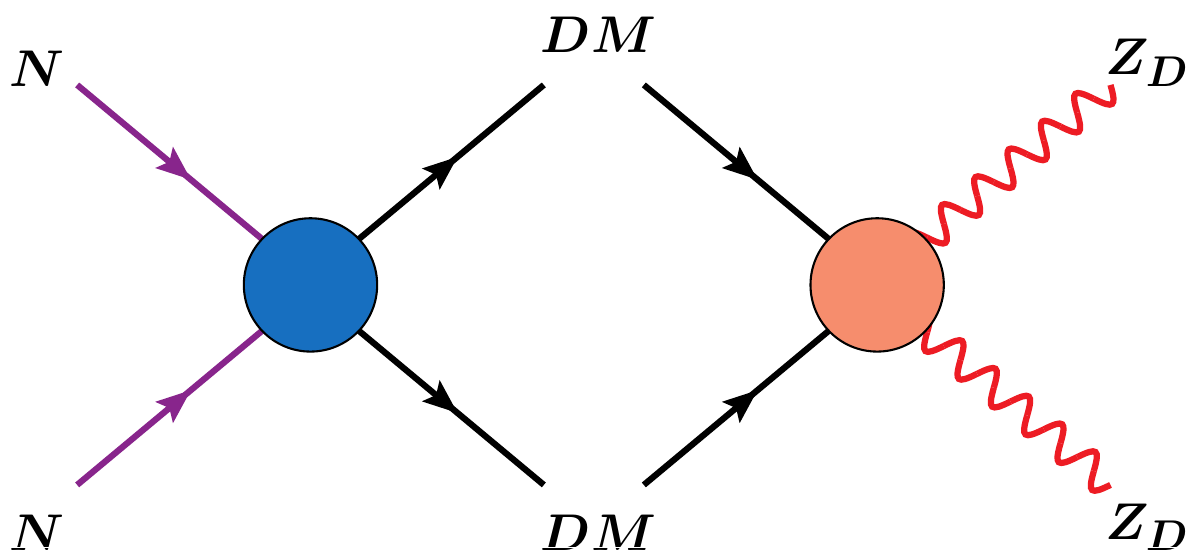}
		\caption{A schematic diagram of the mechanism to generate correct thermal relic of self-interacting DM with a light mediator $(Z_D)$.}
		\label{fig:schematic}
	\end{figure}

	
	 Motivated by this, in this letter, we introduce a minimal scenario for realisation of light thermal SIDM where DM is a Dirac fermion charged under a $U(1)_D$ gauge symmetry. 
	The corresponding gauge boson $Z_D$ is much lighter than DM, leading to the required self-interaction. More specifically, we focus on DM mass in the range: 1-10 GeV, such that DM freeze-out occurs well above the epoch of big bang nucleosynthesis (BBN). This also keeps ZD mass above the limit allowed by BBN constraints (to be discussed later) while being consistent with required DM self-interactions. In such a setup, a large DM annihilation rate into $Z_D$ pairs lead to under-abundant relic after thermal freeze-out, specially for DM in the GeV ballpark. An additional fermion (N) with mass close to but larger than that of DM is introduced which can annihilate into DM efficiently. This production channel compensates for the large depletion of DM relic due to the annihilation of latter into light mediators. A schematic of this mechanism is shown in Fig. \ref{fig:schematic}. The interaction of N with DM is provided by the same singlet scalar responsible for spontaneous breaking of $U(1)_D$ gauge symmetry. In a sharp contrast to the earlier works \cite{Dutta:2021wbn, Borah:2021pet, Borah:2021rbx, Borah:2021qmi}, where non-thermal decay of long-lived particles to DM generates the correct relic, here it is happening largely by the thermal scattering process: ${\rm N \, N \rightarrow DM \, DM}$ occurring around the thermal freeze-out epoch of DM. Here we show that this minimal setup not only leads to correct thermal DM relic and required self-interactions but also remain verifiable at experiments related to dark photon searches, cosmic microwave background (CMB) as well as direct detection experiments sensitive to light DM. Additionally, the singlet fermion (N) can be part of a bigger setup where N and its heavier partners can play a role in generating light neutrino masses via seesaw mechanism. This also opens up an indirect detection prospect of DM in terms of diffused or mono-chromatic photons along with specific predictions for lightest active neutrino mass. Thus, the minimal setup proposed here encompasses several mysteries related to astrophysics, cosmology and particle physics while having the potential of being verified at near future experiments.
	
	We first outline the minimal setup followed by the details of DM self-interaction and relic density. We then discuss the detection prospects of this minimal setup and briefly comment upon the possibility of connecting it to the origin of light neutrino mass via Majorana or Dirac seesaw mechanism of type I. \\

\noindent
{\bf Minimal setup:} We consider DM to be a Dirac fermion $\psi$ having charge 1 under a hidden $U(1)_D$ gauge symmetry. A singlet scalar $\Phi$ having the same charge is responsible for spontaneous breaking of $U(1)_D$ gauge symmetry. An additional singlet fermion N (considered to be of Dirac type for simplicity) with vanishing $U(1)_D$ charge is introduced to realise the required thermal relic of DM. The relevant Lagrangian is given by:
	\begin{eqnarray}
		\mathcal{L} &\supset i \overline{\psi} \gamma^\mu D_\mu \psi - M_{\psi} \overline{\psi} \psi -M_{N} \overline{N} N\nonumber\\&
		-\{ Y_{\psi N}\overline{\psi} \Phi N  + {\rm h.c.} \}+ \frac{\epsilon}{2}B^{\alpha \beta}Y_{\alpha\beta}  
	\end{eqnarray}
	where $D_\mu = \partial_\mu + i g_D (Z_{D})_\mu$ and $B^{\alpha\beta}, Y_{\alpha \beta}$ are the field strength tensors of $U(1)_D, U(1)_Y$ respectively with $\epsilon$ being the kinetic mixing between them. 
 The singlet scalar can acquire a non-zero vacuum expectation value (VEV) $\langle \Phi \rangle =v_\phi$ giving rise to $U(1)_D$ gauge boson mass $M_{Z_{D}}= g_D v_\phi$. The same singlet scalar VEV also generates a mixing between $\psi$ and $N$. In the basis $\left(\psi,N\right)^T$, the mass matrix for the fermions can be written as
$$	\left(
\begin{array}{cc}
M_\psi &Y v_\phi/\sqrt{2} \\
Y v_\phi/\sqrt{2} & M_N \\
\end{array}
\right)\,.
$$
With the diagonalisation of the above mass matrix we get the physical states: $\chi_1 =\cos\theta ~\psi -\sin\theta~ N$ and 
$\chi_2=\sin\theta~ \psi + \cos\theta~ N$, 
where the mixing parameter is given by
\begin{equation}\label{mixangle}
    \tan~2\theta=\frac{\sqrt{2}Y_{\psi N} v_\phi}{M_N-M_\psi}.
\end{equation}
Here $\chi_1$ with mass $M_{\chi_1}$ being the lightest becomes the stable DM candidate and $\chi_2$ with mass $M_{\chi_2}= M_{\chi_1}+\Delta M$ is the next to lightest stable particle (NLSP). Due to the scalar quartic coupling, the neutral scalar part of $H$ and $\Phi$, {\it i.e.} $h$ and $\phi$ respectively, mix with each other via an angle $\beta$.


	\begin{figure}[h!]
		\centering
		\includegraphics[width=8cm,height=7.5cm]{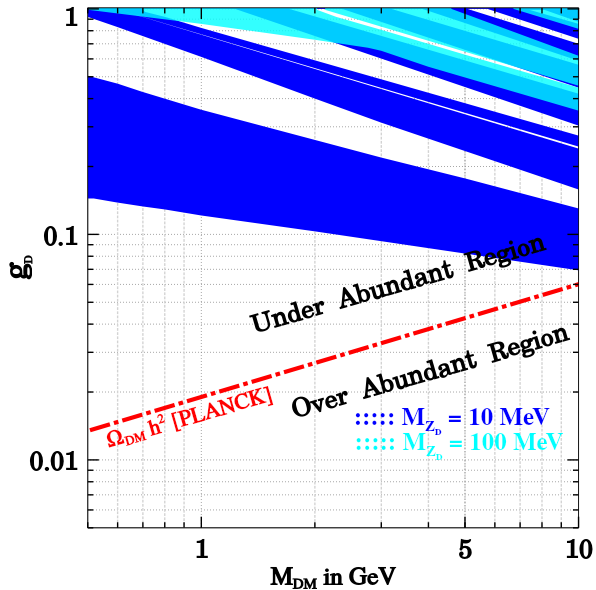}
		\caption{Parameter space in $g_D-M_{\rm DM}$ plane consistent with DM self-interactions (coloured regions) and DM relic (red coloured contour) for $M_{Z_D}=10$ MeV and $100$ MeV with $\cos \theta=1$.}
		\label{fig:sidm_ps_relic_con}
	\end{figure}
   
\noindent 
{\bf DM self-interaction and thermal relic:} Through t-channel processes, the DM $\chi_1$ can have elastic self-scattering via the vector like couplings to $Z_D$ of type $\cos^2\theta g_D (Z_{D})_\mu\overline{\chi_1}\gamma^\mu\chi_1$ which can alleviate the small-scale problems of $\Lambda{\rm CDM}$. The typical DM self-scattering cross-section is many orders of magnitude larger than the cross-section for typical thermal DM and can naturally be realized with the light $Z_D$, much lighter than the typical weak scale mediators. For such light mediator, the self-interaction of non-relativistic DM can be described by a Yukawa type potential: $V(r)= \pm \frac{\alpha_D}{r}e^{-M_{Z_{D}}r}$, where
the + (-) sign denotes repulsive (attractive) potential and $\alpha_D = g^2_D /4\pi$ is the dark fine structure constant. While $\psi \overline{\psi}$ interaction is attractive, $\psi \psi$ and $\overline{\psi} \overline{\psi}$ interactions are repulsive.

	
	To capture the relevant physics of forward scattering divergence, we define the transfer cross-section $\sigma_T=\int d\Omega (1-\cos\theta) \frac{d\sigma}{d\Omega}$~\cite{Feng:2009hw,Tulin:2013teo,Tulin:2017ara}.
	Capturing the whole parameter space requires the calculations to be carried out well beyond the perturbative limit. Depending on the masses of DM ($M_{\chi_1}=M_{\rm DM}$) and the mediator ($M_{Z_D}$) along with relative velocity of DM ($v$) and interaction strength ($g_{D}$), three distinct regimes can be identified, namely the Born regime ($g^2_{D} M_{\rm DM}/(4\pi M_{Z_D}) \ll 1,  M_{\rm DM} v/M_{Z_D} \geq 1$), classical regime ($g^2_{D} M_{\rm DM}/4\pi M_{Z_D}\geq 1$) and the resonant regime ($g^2_{D} M_{\rm DM}/(4\pi M_{Z_D}) \geq 1, M_{\rm DM} v/M_{Z_D} \leq 1$). For details, one may refer to \cite{Feng:2009hw,Tulin:2013teo,Tulin:2017ara,Tulin:2012wi,Khrapak:2003kjw}. In Fig. \ref{fig:sidm_ps_relic_con}, we show the parameter space in $g_D-M_{\rm DM}$ plane which can give rise to the desired DM self-interactions for two different values of mediator mass ($M_{Z_{D}}=10$MeV and $100$ MeV). The red coloured dot-dashed line corresponds to the parameter space consistent with thermal relic of DM if only DM annihilation into $Z_D$ is considered. The cross-section for the $\chi_1 \chi_1 \to Z_D Z_D$ process is given by
	\begin{equation}
		\langle\sigma v\rangle \sim \frac{\pi ~\cos^8\theta~ \alpha^2_D}{M^2_{\chi_1}}.
	\end{equation}
	 Clearly, in the low mass regime, the parameter space satisfying self-interaction criteria gives rise to  under-abundant thermal relic.

\begin{figure}
	\centering
		\includegraphics[scale=0.55]{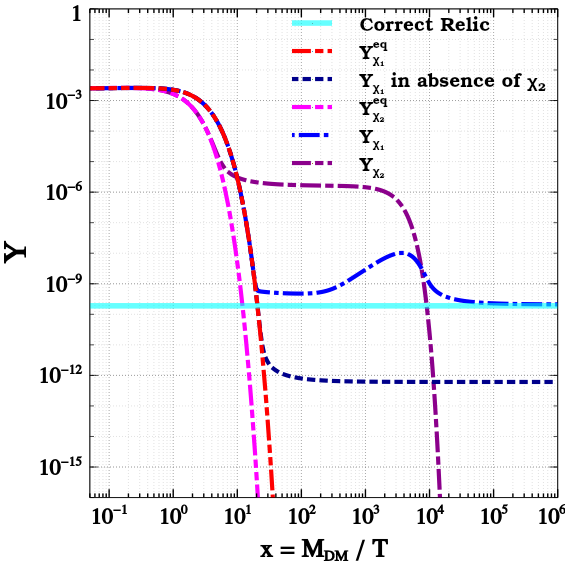}
	\caption{Evolution of comoving number densities of $\chi_1, \chi_2$ for specific benchmark values of the parameters: $M_{\chi_{1}} = 1.685~ {\rm GeV}, \Delta M = 1.19 ~{\rm GeV}, \sin\theta = 1.33\times 10^{-4}, M_{h_{2}} = 3.63 {\rm GeV}, \sin\beta=4.1\times 10^{-4},   g_{D} = 0.1 ~{\rm and}~ M_{Z_{D}} = 10~ {\rm MeV}$.}
	\label{Relic}
\end{figure}

	
	However, in the present setup, due to the presence of NLSP $\chi_2$ with mass not too far above that of $\chi_1$ helps in generating correct DM relic by taking part in several processes affecting the evolution of DM number density. 
It is straightforward to write down the Boltzmann equations, relevant for calculation of DM relic abundance, in terms of respective comoving number densities taking into account of coannihilation effects \cite{Griest:1990kh} as well as decay of NLSP into DM. Here it should be mentioned that, the 2-body decay of $\chi_2$ is kinematically forbidden as $M_{h_2}> M_{\chi_2}$ and only 3-body decay  $\chi_2 \to \chi_1 f \bar{f}$ is possible via $h-\phi$ mixing which is given by $\Gamma_{\chi_2 \rightarrow \chi_1 f \Bar{f}}=\frac{1}{640 \pi^3} \frac{Y^2_{\psi N}}{M^4_{h_2}}\cos^2 2\theta \sin^2 2\beta (\frac{m_f}{v})^2 \Delta M^5$. We solve these coupled equations numerically and show the evolution of different comoving number densities in Fig. \ref{Relic}. Clearly, the DM relic remains under-abundant in absence of the NLSP as shown by the dashed blue line in Fig.~\ref{Relic}. Due to the NLSP induced scattering the final relic of DM satisfies the PLANCK observed value \cite{Planck:2018vyg}.

\begin{figure*}[t]
		\centering
		\begin{tabular}{cc}
			\includegraphics[scale=0.5]{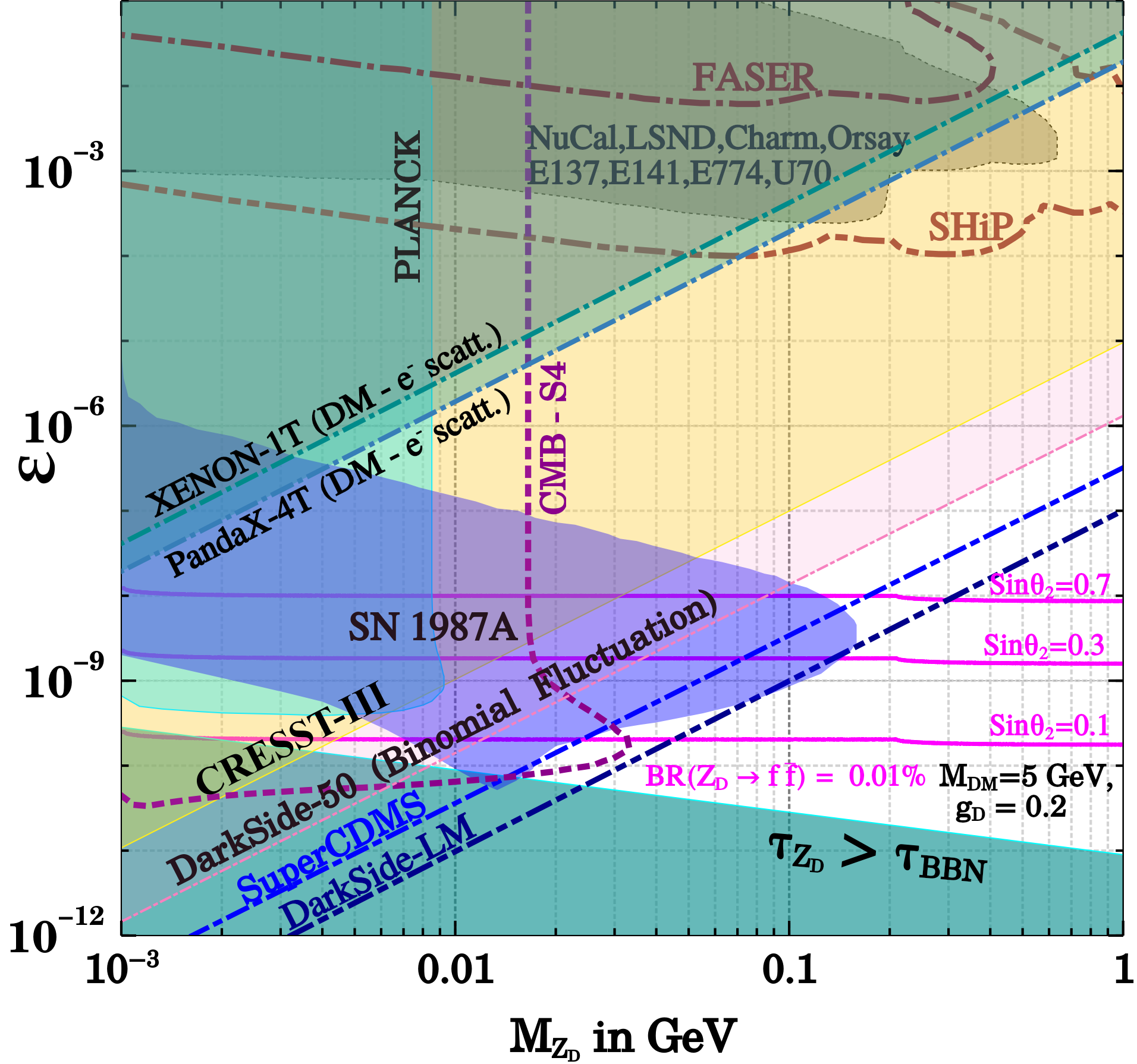}
			\includegraphics[scale=0.5]{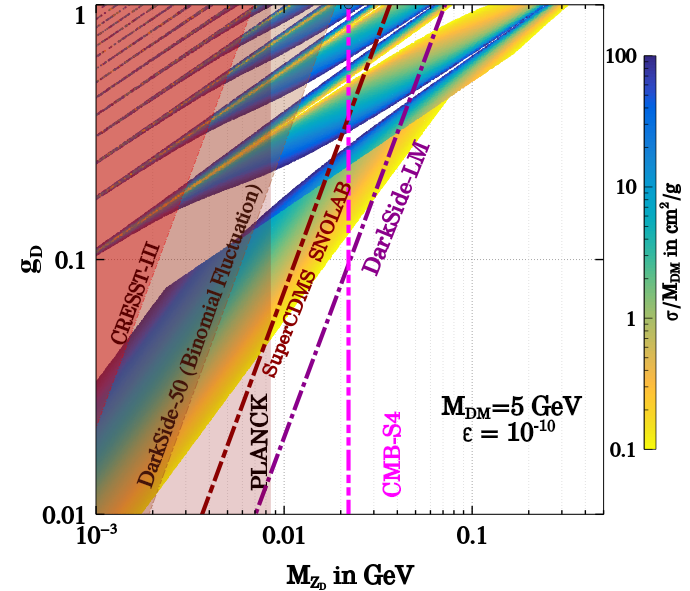}
		\end{tabular}
		\caption{Left panel: summary plot in the plane of kinetic mixing $(\epsilon)$ and $M_{Z_D}$ showing experimental senstivities and constraints for $M_{\rm DM}=5$ GeV, $g_D=0.2$. 
  Right panel: summary plot in the plane of $U(1)_D$ gauge coupling $g_D$ and $M_{Z_D}$ showing DM self-interaction preferred region and experimental constraints, sensitivities for $M_{\rm DM}=5$ GeV, $\epsilon=10^{-10}$. See text for the details.}
		\label{detection}
	\end{figure*}
 
It should be noted that we have considered both $\chi_1$ and $\chi_2$ to be produced in thermal equilibrium. This can happen either due to kinetic mixing ($\epsilon$) of $Z_D$ with $U(1)_Y$ of the SM or due to singlet scalar mixing with the SM Higgs. Since DM self-interactions require sizeable $g_D$ and light $Z_D$, the constraints from dark photon searches as well as astrophysics and cosmology bounds are severe in the $\epsilon-M_{Z_D}$ plane as we show on the left panel of Fig. \ref{detection}. The gray shaded region shows the parameter space excluded by the electron beam dump experiments such as  SLAC-E137, SLAC-E141\cite{Bjorken:2009mm,Andreas:2012mt}, Fermilab-E774\cite{Bross:1989mp}, Orsay\cite{Davier:1989wz}, where the dark gauge boson can be produced in the Bremsstrahlung process. Similarly at proton beam dump facilities, such as CHARM\cite{CHARM:1985nku}, LSND\cite{LSND:1997vqj} and U70/Nu-Cal\cite{Blumlein:2013cua}, hidden photons can be produced in Bremsstrahlung as well as in meson decays produced in proton collisions with the target material. We have also shown the future sensitivity of SHiP facility\cite{SHiP:2015vad} shown by the brown dotted line and the FASER~\cite{Feng:2017uoz}, searching for very displaced hidden photon decays at the LHC, by the maroon coloured dot-dashed line. {However, these constraints are considerably milder in our model compared to the standard scenarios involving a dark boson. In our framework, in order to align with the CMB constraints, the branching fraction of $Z_D$ decay into SM charged fermions is less than $0.01\%$. Consequently, these constraints are relaxed by a factor of $10^{-4}$ compared to conventional searches for a dark photon.
For the same reason, constraints from experiments utilizing fixed targets, such as NA48\cite{NA482:2015wmo}, APEX\cite{APEX:2011dww}, and collider experiments where dark bosons can be generated through s-channel processes and meson decays, such as LHCb\cite{LHCb:2017trq}, as well as $e^+~e^-$ colliders like BaBar\cite{BaBar:2014zli}, Belle\cite{Belle-II:2010dht}, and KLOE\cite{Anastasi:2015qla}, do not exclude any parameter space in Fig.~\ref{detection}. 
Apart from these constraints, the light gauge boson mass and mixing parameter are also constrained from astrophysics and cosmology. Very light $Z_D$ below a few MeV is ruled out from cosmological constraints on effective relativistic degrees of freedom\cite{Planck:2018vyg} which has been shown by the cyan coloured shaded region. This arises due to fact that  late decay of such light gauge bosons into SM leptons, after standard neutrino decoupling temperatures can enhance $N_{\rm eff}$. We also showcase the parameter space sensitive to the future sensitivity of CMB-S4 experiment\cite{Ibe:2019gpv} by the purple dotted line. As a conservative bound, we have also constrained the $Z_D$ lifetime to be less than typical BBN epoch so as not to disturb the predictions of light nuclei abundance by injecting entropy and this has been shown by the green shaded region at the bottom. The constraints from Supernova 1987A\cite{Chang:2016ntp} has been shown by the purple shaded region,  which arises because such dark bosons, produced in sufficient
quantity, can reduce the amount of energy emitted in the form of neutrinos, in conflict with the observation. The constraint from direct detection experiment CRESST-III~\cite{Abdelhameed:2019hmk} and DarkSide-50~\cite{DarkSide:2018bpj} which constrains the DM-nucleon scattering cross-section is shown by the Yellow and Pink shaded regions for a fixed $M_{\rm DM}=5$ GeV and $g_{D}=0.2$. The constraints on the DM-electron scattering rates from Xenon1T~\cite{XENON:2019gfn} and PandaX-4T~\cite{PandaX:2022xqx} has also been shown by green and light blue shaded regions. For the same benchmark values of $M_{\rm DM}$ and $g_{D}$, we have also shown the projected sensitivity of SuperCDMS~\cite{SuperCDMS:2016wui} and DarkSide-LM~\cite{GlobalArgonDarkMatter:2022xgs} by the light and dark blue dot dashed lines respectively.

In addition to all these constraints, the signal of DM annihilation at present can be detected at indirect detection experiments\cite{Fermi-LAT:2015att,HESS:2018cbt,Profumo:2017obk}. Also the annihilation of SIDM to light mediators during recombination can cause distortion in the anisotropy of the CMB as the mediator can decay to electrically charged fermions which puts stringent constraints\cite{Madhavacheril:2013cna,Slatyer:2015jla,Planck:2018vyg, Elor:2015bho, Profumo:2017obk}. 
However, we note that these bounds can be evaded if the mediator dominantly decays into neutrinos or some dark radiations. In our setup, the presence of a heavier dark fermion $(\Psi_2)$ mixing maximally with heavier cousins of N ($N_2$ with mass $\mathcal{O}(100)$ GeV), which are naturally present playing non-trivial role in neutrino mass generation via seesaw mechanism, can facilitate the dominant decay of $Z_D$ into neutrinos. Such decay of $Z_D$ into neutrinos are governed by $\Psi_2-N_{2}$ mixing ($\sin\theta_2$)  and $\nu-N_2$ mixing ($\sin\theta_{\nu N_2}$) whereas the decay of $Z_D$ into charged fermions are governed by the kinetic mixing $\epsilon$. 
{The corresponding decay widths are estimated to be
\begin{eqnarray}
    \Gamma^{Z_D}_{ f\bar{f}}&=& \frac{\epsilon^2 g^2 M_{Z_{D}}}{12 \pi} \left(1+\frac{2 m^2_f}{M^2_{Z_{D}}}\right)\left(1-\frac{4 m^2_f}{M^2_{Z_{D}}}\right)^{1/2},\nonumber\\
    \Gamma^{Z_D}_{\nu \bar{\nu}}&\simeq&\frac{ g^2_D \sin^4\theta_2\sin^4\theta_{\nu N_2}M_{Z_{D}}}{12 \pi}  \left(1+\frac{2 m^2_\nu}{M^2_{Z_{D}}}\right)\left(1-\frac{4 m^2_\nu}{M^2_{Z_{D}}}\right)^{1/2}.\nonumber\\
\end{eqnarray}
We have imposed a conservative upper limit on the branching ratio of $Z_D$ into charged leptons, setting it at $0.01\%$, in order to establish the upper boundary on the kinetic mixing parameter. This limit is illustrated in the left panel of Fig.~\ref{detection} by the solid magenta contours, considering three different values of the $\sin\theta_2$ mixing for a typical $\sin\theta_{\nu N_2}=10^{-3}$, which is in line with data on light neutrinos.
It is noteworthy that this cautious upper bound incorporates three possible final states for dark matter annihilation during the recombination epoch: I) $2$ SM charged fermion pairs, II) $1$ charged fermion and $1$ neutrino pair, and III) $4$ neutrino final states. It is evident that the annihilation rate to charged fermions in case-II surpasses that in case-I when the branching of $Z_D\to f\bar{f}$ is very small. Therefore, we employ this channel to constrain our parameter space, as it provides the most rigorous constraint.} The region above these contours is ruled out from CMB and indirect detection constraints for that particular value of $\sin\theta_2$.

In the right panel of Fig.~\ref{detection}, we have confronted the $g_D$ and $M_{Z_D}$ parameter space satisfying correct self-interaction criteria against various constraints from direct search of DM as well as cosmology. The coloured shaded region depicts the parameter space that gives rise to correct self-interaction cross-section for a fixed $M_{\rm DM}=5$ GeV with the colour-bar depicting the value of the $\sigma /M_{\rm DM}$ in units of ${\rm cm}^2/g$. The present constraints from PLANCK on $N_{\rm eff}$ and CRESST-III constraint on DM-nucleon scattering rules out very small values of $M_{Z_{D}}<8.5$ MeV. And interestingly, the correct self-interaction parameter space comes within the projected sensitivity of CMB-S4\cite{Ibe:2019gpv} as well as direct search experiments which are shown by the dotted lines\cite{SuperCDMS:2016wui,GlobalArgonDarkMatter:2022xgs} for a fixed value of $\epsilon=10^{-10}$ which is consistent with all relevant constraints shown in the left panel of Fig.~\ref{detection}. While we have chosen a benchmark $M_{\rm DM}=5$ GeV in the summary plots, choosing a different value like 1 GeV or 10 GeV slightly shifts the shaded regions and contours without changing the general conclusions inferred from Fig. \ref{detection}.

\noindent
{\bf Connection to neutrino mass:} If $N$ and its heavier cousins are of Dirac nature, as has been assumed in the above discussions, they can give rise to the light Dirac neutrino masses via type I seesaw mechanism \cite{Borah:2017dmk}. To implement this, we need to include the right chiral parts of Dirac neutrinos namely, $\nu_R$ and another singlet scalar $(\eta)$ both of which are odd under a softly broken $Z_2$ symmetry and neutral under the SM and $U(1)_D$ gauge symmetries. Similar to typical Dirac neutrino models, a global lepton number is assumed to be present in order to prevent the Majorana mass terms of singlet fermions. The relevant Yukawa Lagrangian is then given by
\begin{equation}
    -\mathcal{L}_{\rm Yukawa} \supset Y_D \overline{L} \tilde{H} N_R + Y'_D \overline{N_L} \nu_R \eta + {\rm h.c.}
\end{equation}
After $\eta$ acquires a non-zero VEV by virtue of the soft-breaking term $\mu_{\eta H} (\eta H^\dagger H)$, the light neutrino mass can be estimated to be 
\begin{equation}
    m_\nu = -Y'_D M^{-1}_N Y_D \langle H^0 \rangle \langle \eta \rangle
\end{equation}
On the other hand, if N is of Majorana type, then the conventional type I seesaw \cite{Minkowski:1977sc, GellMann:1980vs, Mohapatra:1979ia, Schechter:1980gr, Schechter:1981cv} follows without any need of introducing additional fields. However, in this case, the DM calculation will change as DM will be a Majorana fermion due to mixing with N resulting in inelastic coupling of DM with $Z_D$ \cite{Schutz:2014nka, Blennow:2016gde, Zhang:2016dck, Alvarez:2019nwt, Dutta:2021wbn}. Nevertheless, the desired thermal relic and self-interactions can still be obtained and hence our generic conclusions remain valid in this scenario as well.

Irrespective of Dirac or Majorana nature of N and hence DM, such connection to the type I seesaw mechanism also opens up DM decay modes into SM particles. For DM in the GeV ballpark, the most distinctive decay is the monochromatic gamma-ray line at $E=M_{\rm DM}/2$ generated via one-loop decay $\chi_1 \rightarrow \gamma \nu$. The corresponding decay width can be estimated as \cite{Pal:1981rm, Shrock:1982sc}
\begin{equation}
    \Gamma_{\rm DM} \approx \frac{9 \alpha G^2_F}{256 \pi^4} \sin^2{\theta} \sin^2{\theta_{N\nu}} M^5_{\rm DM}
\end{equation}
where $\theta_{N \nu}$ is the mixing between $N$ and active neutrinos, $\alpha=1/137$ is the fine structure constant and $G_F=1.166 \times 10^{-5} \, {\rm GeV}^{-2}$ is the Fermi constant. From gamma-ray searches, constraints on this decay width comes out to be $\Gamma^{-1}_{\rm DM} \gsim 10^{29}$ s for $M_{\rm DM} = 10$ GeV \cite{Fermi-LAT:2013thd}. Considering $M_{\rm DM} \sim 10$ GeV and $\sin{\theta} \sim 10^{-2}$, we get $\sin{\theta_{N\nu}} \lsim 10^{-20}$. This leads to an almost vanishing lightest active neutrino mass from seesaw mechanism. This will keep the effective neutrino mass much out of reach from ongoing tritium beta decay experiments like KATRIN \cite{KATRIN:2019yun} so that any positive results from this experiment in future can falsify our scenario. Additionally, near future observation of neutrinoless double beta decay \cite{Dolinski:2019nrj} can also falsify our scenario, particularly for normal ordering of light neutrinos as such experiments can probe normal ordering only for $m_{\rm lightest} > 10^{-2}$ eV which lies much above the tiny value predicted in our scenario. \\

\noindent
{\bf Conclusion:} We have proposed a novel and minimal scenario to realise thermal relic of light DM in GeV ballpark with large self-interactions required to solve the small-scale structure problems associated with CDM paradigm. While the presence of a light mediator leads to the required velocity dependent DM self-interactions, the same interactions also lead to under-abundant thermal relic in the GeV mass regime. Assuming DM to be a Dirac fermion with a hidden $U(1)_D$ gauge boson $Z_D$ as light mediator, we consider the presence of another fermion (N), close to but heavier than that of DM mass such that efficient annihilation of N into DM can compensate for large dilution of DM due to its annihilation into $Z_D$ pairs. We show that correct thermal relic can be obtained in such a setup after incorporating all relevant constraints. { Another novelty of this scenario is that it paves a way to generate neutrino mass through 
seesaw mechanism while keeping relic density parameter space intact.} Future experiments related to dark photon search, CMB, direct and indirect detection as well as neutrino physics can play vital role in probing most part of the parameter space for GeV scale self-interacting DM.

\acknowledgements
The work of NS is supported by Department of Atomic Energy-Board of Research in Nuclear 
Sciences, Government of India (Ref. Number: 58/14/15/2021- BRNS/37220).

	\twocolumngrid

\end{document}